\documentclass[twocolumn,english,aps]{revtex4}
\usepackage[T1]{fontenc}
\usepackage[latin9]{inputenc}
\usepackage{color}
\usepackage{babel}
\usepackage{amsmath}
\usepackage{amssymb}
\usepackage{graphicx}
\usepackage{esint}
\usepackage[unicode=true,pdfusetitle,
 bookmarks=true,bookmarksnumbered=false,bookmarksopen=false,
 breaklinks=false,pdfborder={0 0 1},backref=false,colorlinks=true]
 {hyperref}

\makeatletter

\providecommand{\tabularnewline}{\\}

\@ifundefined{textcolor}{}
{%
 \definecolor{BLACK}{gray}{0}
 \definecolor{WHITE}{gray}{1}
 \definecolor{RED}{rgb}{1,0,0}
 \definecolor{GREEN}{rgb}{0,1,0}
 \definecolor{BLUE}{rgb}{0,0,1}
 \definecolor{CYAN}{cmyk}{1,0,0,0}
 \definecolor{MAGENTA}{cmyk}{0,1,0,0}
 \definecolor{YELLOW}{cmyk}{0,0,1,0}
}

\makeatother

\begin{document}

\title{Predicting the Influence of Plate Geometry on the Eddy Current Pendulum}

\author{Catherine Weigel}

\affiliation{Department of Physics and Astronomy, Tufts University, Medford, MA
02155, USA }

\author{Jeremy Wachter}

\affiliation{Department of Physics and Astronomy, Tufts University, Medford, MA
02155, USA }

\author{Paul Wagoner}

\affiliation{Department of Physics and Astronomy, Tufts University, Medford, MA
02155, USA }

\author{Timothy J. Atherton}

\email{timothy.atherton@tufts.edu}

\selectlanguage{english}%

\affiliation{Department of Physics and Astronomy, Tufts University, Medford, MA
02155, USA }
\begin{abstract}
We quantitatively analyze a familiar classroom demonstration, Van
Waltenhofen's eddy current pendulum, to predict the damping effect
for a variety of plate geometries from first principles. Results from
conformal mapping, finite element simulations and a simplified model
suitable for introductory classes are compared with experiments. 
\end{abstract}
\maketitle

\section{Introduction}

\begin{figure}
\includegraphics{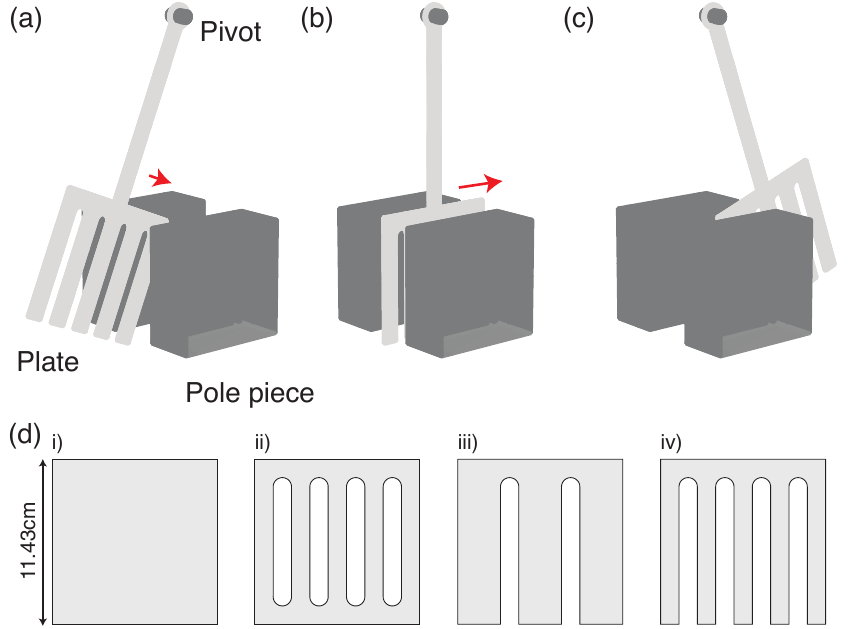}

\protect\caption{\label{fig:schematic}Schematic of Von Waltenhofen's pendulum. (a)
An aluminium plate swings between the pole pieces of an electromagnet.
As it passes through the fringing field, eddy currents are induced
in the plate that dissipate energy and damp the pendulum. (b) As the
plate passes through the center of the pole pieces, the rate of change
in flux through the plate vanishes and so the eddy current momentarily
vanishes. (c) As the pendulum swings back, the eddy current reverses
direction. (d) Various plate geometries under investigation in the
present work (to scale).}
\end{figure}

Eddy currents are induced electric currents in a conducting material
that result when either the object moves through a nonuniform magnetic
field or is stationary but subject to a time-changing magnetic field.
They were first observed by Fran\c{c}ois Arago\citep{Babbage01011825}
but it was not until Michael Faraday's discovery of induction\citep{Faraday01011832}
that the mechanism of this phenomenon was understood. A pioneering
experiment to quantify the heat dissipated by the eddy currents, and
hence connect electromagnetism to thermodynamics, is von Waltenhofen's
pendulum\citep{vonWaltenhofen:1879p4081} {[}figure \ref{fig:schematic}(a)\textemdash (c){]}.
The bob of the pendulum is a non ferromagnetic conducting plate which
swings between the poles of an electromagnet, generating circulating
eddy currents. The dissipative eddy currents tend to oppose the motion
of the bob, and hence damp the pendulum. The experiment is an example
of magnetic braking: when the magnet is turned off, the pendulum oscillates
as a classic pendulum, damped only by air resistance and frictional
forces in the hinge; when the magnetic field is switched on, the pendulum
comes to a rapid stop. 

Von Waltenhofen's pendulum is a popular demonstration in high school\citep{Thompson:2011p4082}
and introductory undergraduate\citep{Onorato:2012p4083} classes because
it provides a striking visual illustration of the invisible eddy currents.
It is easy to set up, and gives the instructor a chance to discuss
real-world applications such as magnetic braking in roller coasters,
hybrid cars and the detection of flaws in conductive materials. Moreover,
by using different shaped plates as bobs, some of the characteristics
of eddy currents can be explored: if the plate is now substituted
with one where the slits do not reach the edges {[}figure 1(d)(ii){]},
the damping is only mildly altered. If a solid plate is replaced by
a comb-like plate with slits cut into it {[}figure 1(d)(iii)-(iv){]}
the damping effect is greatly attenuated. Students infer from this
that the eddy currents must be localized to some region of the plate. 

Typically, however, the relationship between the shape of the plate
and the damping is considered only qualitatively due to the complexity
of the calculations required, which require advanced mathematical
techniques introduced in higher-level electromagnetism classes. In
this paper, therefore, we apply some of these strategies to predict
the damping behavior of conducting plates with different geometries.
We investigate four plate shapes, illustrated in figure 1(d): (i)
a square plate, (ii) a plate with holes that do not reach the edge,
(iii) a plate with with two slots, (iv) another with four slots. The
present work was primarily performed during a project-based graduate
electrodynamics class with the objective of giving students a real,
tractable example of the utility of their calculations. It was also
an opportunity to address the challenges of connecting theory to experiment,
even where the fundamental theory is uncontroversial. 

The paper is organized as follows: in section II, the form of the
damping for the present experiment is derived first in terms of a
simplified model suitable for undergraduate classes and the effect
of these damping terms on the simple pendulum is explored. In section
III, we present a more sophisticated model derived from Maxwell's
equations in the magneto-quasistatic approximation. Within this framework,
the rate of power dissipation for each plate is estimated using a
conformal mapping technique and a finite element solution. In section
IV, the motion of the pendulum with each of the four plates is determined
experimentally and compared with the predictions of each of the theoretical
models in section V. In section VI, the pedagogical context of this
work is discussed. Brief conclusions are presented in section VII. 

\section{Simplified Model\label{sec:Simplified-Model}}

The damped simple pendulum is, in the presence of dissipative forces,
described by the equation,
\begin{equation}
\ddot{q}+\omega_{0}^{2}q+\frac{\partial D}{\partial\dot{q}}=0\label{eq:pendulumeqn}
\end{equation}
where $D(q,\dot{q})$ is the Rayleigh dissipation function, $\omega_{0}$
is the natural frequency of the pendulum and $q(t)$ is a generalized
coordinate. In the absence of the magnetic field, a Stokes-like drag
is assumed,
\begin{equation}
D_{stokes}=\beta\dot{q}^{2},
\end{equation}
arising from friction in the bearing and air resistance. In this section
and the next, the form of the dissipation function in the presence
of the magnetic field is derived, first for a simplified model suitable
for introductory classes and second a more physically correct model
based on the magneto-quasistatic approximation. The remaining theory
section considers the effect of these terms on the solutions of (\ref{eq:pendulumeqn}).

To develop the simplified model, we make the gross approximation that
the eddy currents are completely localized to the boundary of the
plate. Then, the plate may be replaced by a thin closed loop of wire
of equivalent shape with area $A$, length $L$, resistivity $\rho$
and with a normal unit vector $\hat{\mathbf{n}}$. The magnetic flux
passing through the loop is,
\begin{equation}
\Phi=\int\mathbf{B}\cdot d\mathbf{A}=\left\langle \mathbf{B}\right\rangle A
\end{equation}
where $\left\langle \mathbf{B}\right\rangle $ is the mean value of
the magnetic field over the plate when it is located at a particular
position $q$. The emf induced around the loop is, by the Faraday-Lenz
law and using the chain rule,
\begin{equation}
V=-\frac{\partial\Phi}{\partial t}=-\frac{\partial\Phi}{\partial x}\dot{q}=-A\frac{d\left\langle \mathbf{B}\cdot\hat{\mathbf{n}}\right\rangle }{dq}\dot{q}.
\end{equation}
The power dissipated in the loop is precisely the dissipation function
desired and readily calculated,
\begin{equation}
D=V^{2}/R,
\end{equation}
which, using $R=\rho L$, becomes,
\begin{equation}
D=\underbrace{\left(\frac{A^{2}}{L}\right)}\frac{1}{\rho}\left(\frac{d\left\langle \mathbf{B}\cdot\hat{\mathbf{n}}\right\rangle }{dq}\right)^{2}\dot{q}^{2}\label{eq:RayleighDissipationFunction}
\end{equation}
where the highlighted term represents the ``geometric factor'' arising
from the shape of the plate. The dissipation function (\ref{eq:RayleighDissipationFunction})
has the form of a position-dependent Stokes drag. 

It is now necessary to solve the pendulum equation including the new
dissipation terms (\ref{eq:RayleighDissipationFunction}),
\begin{equation}
\ddot{q}+\omega_{0}^{2}q+2\left[\beta+\frac{A^{2}}{L\rho}\left(\frac{d\left\langle \mathbf{B}\cdot\hat{\mathbf{n}}\right\rangle }{dq}\right)^{2}\right]\dot{q}=0,
\end{equation}
for which the form of the magnetic field must be known. Here, the
coordinate system is oriented as shown in figure 1 such that the plate
normal $\hat{\mathbf{n}}$ is parallel to the $z$ axis, so only $B_{z}$
is required; the origin is chosen to lie in the middle of the pole
pieces. An approximation made in other studies\citep{Thompson:2011p4082,Onorato:2012p4083}
is that the magnetic field is constant between the pole pieces and
zero everywhere else, i.e. that,
\begin{equation}
B_{z}=\begin{cases}
B_{0} & -L_{pp}/2\leq x<L_{pp}/2\\
0 & x<-L_{pp}/2,\ x\ge L_{pp}/2
\end{cases}\label{eq:bansatz}
\end{equation}
where $L_{pp}$ is the width of the pole pieces. If the plate is square
of side $L$, the average value of the field over the surface of the
plate when it is centered at $q$ along the $x$-axis may be determined,
\begin{equation}
\left\langle \mathbf{B}\cdot\hat{\mathbf{n}}\right\rangle =L\int_{q-L/2}^{q+L/2}B_{z}(x')\ dx'.
\end{equation}
For the ansatz (\ref{eq:bansatz}), the flux through the plate increases
(decreases) linearly as it enters (leaves) the field. Depending on
the size of the plate, and the amplitude of the motion, the plate
may also experience portions of the motion where there is no change
in flux. If the plate is the same size as the pole pieces $L=L_{pp}$,
and never fully leaves the field, i.e. $\left|x\right|<(L_{pp}+L)/2$
, then $\left\langle \mathbf{B}\cdot\hat{\mathbf{n}}\right\rangle \sim B_{0}q$
and the classical damped pendulum is recovered, but with a modified
damping constant, 
\begin{equation}
\ddot{x}+\omega_{0}^{2}x+2\left[\beta+\frac{A^{2}B_{0}^{2}}{L\rho}\right]\dot{x}=0.\label{eq:equationofmotion}
\end{equation}

\begin{figure}
\begin{centering}
\includegraphics{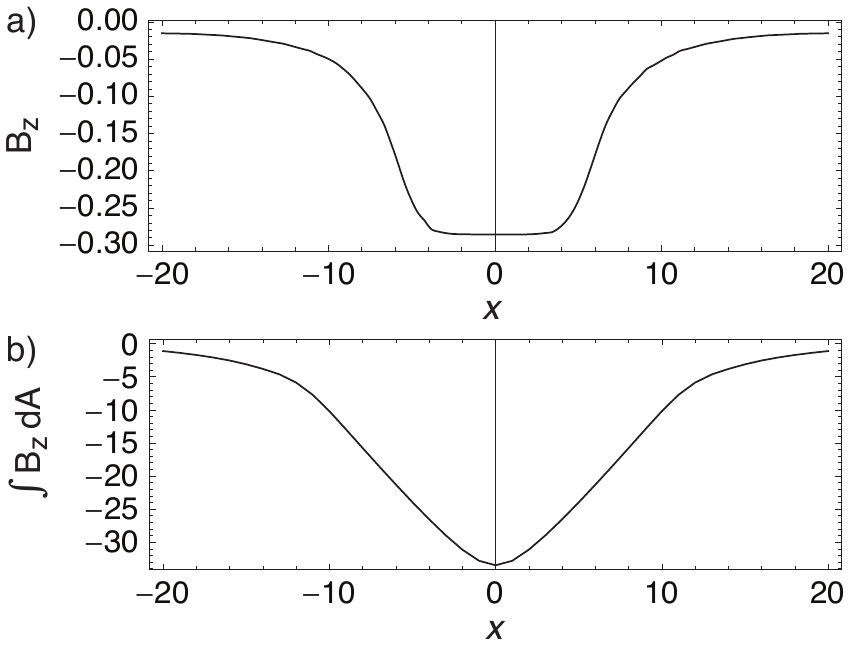}
\par\end{centering}

\protect\caption{\label{fig:FieldProfile}(a) Field strength as a function of distance
$x$ through the center of the pole pieces. (b) Magnetic flux through
a square plate as a function of the position of the plate $x$.}
\end{figure}

In order to gauge the validity of the (\ref{eq:bansatz}) for the
pendulum we calculated the spatial dependence of the magnetic field
due to the pole pieces in the magnetostatic formulation, using a scalar
potential function defined $\mathbf{B}=-\nabla\phi$ and subject to
Laplace's equation $\nabla^{2}\phi=0$. This was solved using the
finite element software \texttt{FlexPDE} for square pole pieces of
width $11$cm and separation $3.5$cm. Dirichlet boundary conditions
were imposed on the pole pieces. A plot of $B_{z}$ along the horizontal
centerline of the midplane of the pole pieces is displayed in fig.
\ref{fig:FieldProfile}(a), showing as expected that the field is
constant between the pole pieces with the fringing field decaying
rapidly outside. From this calculated profile, the flux $\int\mathbf{B}\cdot d\mathbf{A}$
through a square plate placed at different horizontal positions was
also calculated; the result is displayed in fig. \ref{fig:FieldProfile}(b).
Notice that the flux depends linearly on position for $1\lesssim\left|x\right|\lesssim10$cm,
suggesting that the equation of motion (\ref{eq:equationofmotion})
is likely to be valid for moderate amplitudes; however around $x=0$
the flux depends quadratically on $x$, suggesting that (\ref{eq:equationofmotion})
is likely to be inaccurate for small amplitudes. 

\section{Magneto-quasistatic Model\label{sec:Magneto-quasistatic-Model}}

Since the eddy current is not strictly confined to the edges of the
plate, it is necessary to develop a more sophisticated model by solving
Maxwell's equations 
\begin{equation}
\nabla\cdot\mathbf{E}=0,\ \nabla\cdot\mathbf{B}=0,\ \nabla\times\mathbf{H}=\mathbf{J},\ \nabla\times\mathbf{E}+\frac{\partial\mathbf{B}}{\partial t}=0\label{eq:Maxwell}
\end{equation}
together with the continuity equation
\begin{equation}
\nabla\cdot\mathbf{J}+\frac{\partial\rho}{\partial t}=0\label{eq:continuityequation}
\end{equation}
to determine the true current density in the plate. Here, the \emph{magneto-quasistatic}
approximation\citep{jackson1975classical} has been made, i.e. the
Maxwell correction to Ampere's law is neglected. This is justified
because the timescale of variation of the magnetic field over the
surface of the plate is sufficiently long that the eddy current produced
does not significantly alter the magnetic field in turn. 

To perform the analysis, consider a thin metal plate of uniform conductivity
$\sigma$ and thickness $d$ and where the shape of the plate is arbitrary.
It is assumed that Ohm's law holds for the plate, so
\begin{equation}
\mathbf{J}=\sigma\mathbf{E}.\label{eq:ohm}
\end{equation}
In the frame of the laboratory, since the pendulum is rigid, the position
of the plate may be described by the single generalized coordinate
$q(t)$ in eq. \ref{eq:pendulumeqn}. As the pendulum swings, the
plate passes through regions of different magnetic field strength.
Viewed from the point of view of the plate, the magnetic field across
its surface changes as a function of time. This may be described by
defining a local coordinate system $(x',y',z')$ relative to the plate
such that the plate lies in the $x'-y'$ plane, and considering the
magnetic field in this frame, 
\begin{equation}
\mathbf{B}=\mathbf{B}(x',y',q(t)).
\end{equation}
We note that the reference frame $(x',y',z')$ is \emph{not} an inertial
frame as the pendulum undergoes periodic acceleration during the course
of its motion; however we neglect these effects as the pendulum velocity
$r\dot{q}\ll c$. By Faraday's law, an electric field and hence an
eddy current is induced in the plate: if $d\ll\delta$ the skin depth
of the material it may be assumed that $\mathbf{E}$ is oriented in
the $x-y$ plane and does not vary significantly over the thickness
of the plate; hence the problem is quasi-two-dimensional. 

From equations (\ref{eq:Maxwell}) and (\ref{eq:ohm}), it is a standard
textbook problem to show that the quantities $\mathbf{E}$, $\mathbf{B}$,
$\mathbf{A}$ and $\mathbf{J}$ all obey the diffusion equation,
\begin{equation}
\nabla^{2}\mathbf{J}=\sigma\frac{\partial\mathbf{J}}{\partial t}.\label{eq:diffusioneq}
\end{equation}
Since $\rho=0$, the continuity equation implies that $\mathbf{J}$
must have no divergence and hence can be written using a stream function,\emph{
\begin{equation}
J_{x}=\frac{\partial\phi(x',y')}{\partial y},\ J_{y}=-\frac{\partial\phi(x',y')}{\partial x'}.\label{eq:Jstreamfunction}
\end{equation}
}By inserting (\ref{eq:Jstreamfunction}) into (\ref{eq:diffusioneq}),
together with the given applied Magnetic field into Maxwell's equations,
it may be shown that the stream function obeys Poisson's equation
\begin{equation}
\nabla^{2}\phi(x',y')=\sigma\frac{d\mathbf{B}\cdot\hat{\mathbf{n}}}{dq}\dot{q}=B_{0}\sigma\dot{q}\zeta(\mathbf{x}',q)\label{eq:Poisson}
\end{equation}
where $\nabla$ here is the 2D Laplacian, $B_{0}$ is the magnetic
field strength and we defined a source term $\zeta(\mathbf{x},q)=\frac{d\mathbf{B}\cdot\hat{\mathbf{n}}}{dq}/B_{0}$.
Since $\mathbf{J}$ must lie tangentially to the edge of the plate,
$\phi$ obeys a Dirichlet boundary condition $\phi=0$ on all edges
of the plate. From the solution for $\phi$, the instantaneous rate
of power dissipation by the plate is then calculated,
\begin{equation}
D=\frac{d}{\sigma}\int\left|\mathbf{J}\right|^{2}\ \text{d}A,\label{eq:powerdissipation}
\end{equation}
where the integral is to be taken over the plate. Inserting the stream
function (\ref{eq:Jstreamfunction}) into (\ref{eq:powerdissipation}),
integrating by parts and using (\ref{eq:Poisson}), we obtain

\begin{eqnarray}
D & = & \frac{d}{\sigma}\int\nabla\phi\cdot\nabla\phi\ \text{d}A,\label{eq:powerdissipation-1}\\
 & = & \frac{d}{\sigma}\left(\oint\phi\nabla\phi dl-\int\phi\nabla^{2}\phi\ dA\right)\\
 & = & -d\dot{q}B_{0}\int\phi\zeta(\mathbf{x}',q)\ dA.\label{eq:powerdissipationintermediate}
\end{eqnarray}
A formal solution to (\ref{eq:Poisson}) may be constructed using
the Green function for the Laplacian,
\begin{equation}
\phi(\mathbf{x}')=\sigma\dot{q}B_{0}G\zeta=\sigma\dot{q}B_{0}\int G(\mathbf{x}',\mathbf{x}'')\zeta(\mathbf{x}'',q)\ dA''.\label{eq:formalsolution}
\end{equation}
Inserting this into (\ref{eq:powerdissipationintermediate}) and rearranging
we obtain the dissipation function,

\begin{widetext}
\begin{eqnarray}
D(q,\dot{q}) & = & -dB_{0}\dot{q}\int\left[\sigma B_{0}\dot{q}\int G(\mathbf{x}',\mathbf{x}'')\zeta(\mathbf{x}'',q)\ dA''\right]\zeta(\mathbf{x}')\ dA'\\
 & = & -d\sigma B_{0}^{2}\dot{q}^{2}\int\int\zeta(\mathbf{x}',q)\zeta(\mathbf{x}'',q)G(\mathbf{x}',\mathbf{x}'')\ dA'\ dA''.\label{eq:powerdissipation-2}
\end{eqnarray}
\end{widetext}Since the source function $\zeta(\mathbf{x},q)$ depends
on the position of the pendulum $q$, this has the form
\begin{equation}
D(q,\dot{q})\propto B_{0}^{2}P(q)\dot{q}^{2}
\end{equation}
where we define a local dissipation rate $P(q)$. To determine an
effective damping coefficient, the calculations proceed as follows:
in subsection \ref{sub:Evaluating-the-Fringe} the form of $\zeta(\mathbf{x},q)$
is determined, then the positional dependence of the drag term is
evaluated using a conformal mapping technique in subsection \ref{sub:Conformal-Mapping-Solution},
and also by solving (\ref{eq:Poisson}) with \texttt{FlexPDE} in subsection
\ref{sub:Finite-Element-Simulations}; finally, the effect of the
position dependent damping on the motion of the pendulum is considered
in the subsection \ref{sub:Effect-on-the}.

\subsection{Evaluating the Fringe field \label{sub:Evaluating-the-Fringe}}

Before evaluating the power dissipation, it is first necessary to
determine the functional form of the magnetic fringe field that the
plate experiences as it moves between the pole pieces and hence the
function $\zeta(x)$. This could also be obtained from a finite element
calculation as was done for the simplified model in section (\ref{sec:Simplified-Model}).
We made the approximation that only the fringe fields from the sides
of the pole pieces are significant in the damping, and the vertical
extent of the pole pieces are neglected; the formulation above remains
sufficiently general that these details could easily be included in
a more sophisticated calculation of $\zeta(x)$. 

\begin{figure}
\begin{centering}
\includegraphics[width=3in]{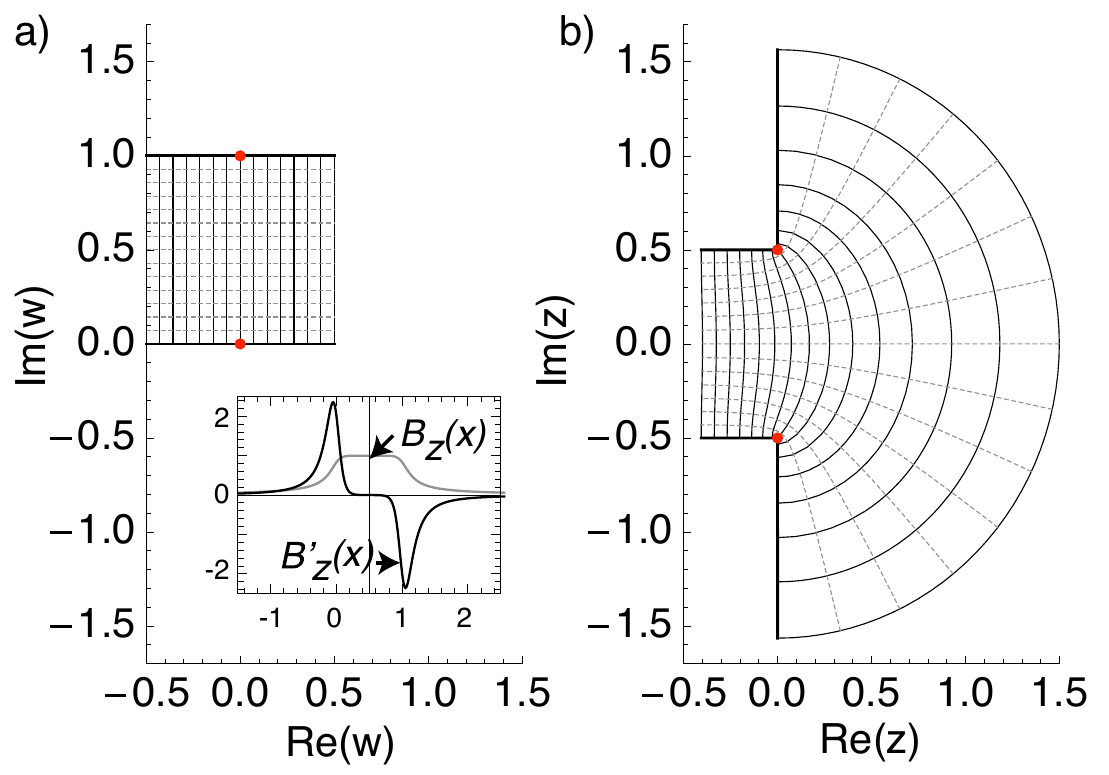}
\par\end{centering}

\protect\caption{\label{fig:CMFringe}Conformal mapping $z(w)$ from (a) a portion
of the strip $0\le\Im(w)\le1$ to (b) the depicted region in the $z$
plane. Places where the integrand in (\ref{eq:scfringe}) vanishes
are indicated by red circles; these introduce corners into the map.
Contours of constant $\Re(w)$ (dashed lines) and $\Im(w)$ (solid
lines) and their projection are indicated corresponding to equipotentials
and field lines respectively. Inset: Plots of the field profile $B_{z}(x)$
and its spatial derivative $B_{z}'(x)$ calculated from the conformal
mapping. }

\end{figure}

Within this approximation, the problem is two dimensional and hence
can be solved using conformal mapping, an important technique for
solving electrostatics problems. For a full discussion we refer the
reader to \citep{visualcomplexanalysis} and \citep{driscoll2002schwarz}.
Briefly, however: A map from a domain to its image is called \emph{conformal}
if angles measured locally around a point in the source domain are
preserved under the mapping. A key property of such maps is that if
a function that is a solution to Laplace's equation is constructed
on some domain, then the projection of that function under the action
of the mapping will \emph{also} obey Laplace's equation. Boundary
value electrostatics problems in two dimensions can hence often be
solved using this method by constructing a solution to Laplace's equation
with appropriate boundary conditions in some simple domain and then
finding a conformal map to the domain of interest. The powerful Riemann
mapping theorem guarantees such a map exists if the domains are simply
connected; mappings between multiply-connected domains may also exist
if certain compatibility criteria are met\citep{driscoll2002schwarz}.
A rich source of such maps are analytic complex functions $z(w)$
that map a domain in the complex plane $w=u+iv$ to an image in the
plane $z=x(u,v)+iy(u,v)$. The required mapping to find the fringe
field is, 
\begin{equation}
z(w)=\int_{0}^{w}\exp(-\pi\xi/2)\left[\sinh\left(\pi\xi\right)\right]^{1/2}\ d\xi.\label{eq:scfringe}
\end{equation}
As shown in figure \ref{fig:CMFringe}, this maps the strip defined
by $0\leq\Im(w)\leq1$ in the $w=u+iv$ plane onto the depicted region
in the $z=x+iy$ plane; the resulting shape resembles an overhead
view of the edge of two parallel pole pieces. The integrand in (\ref{eq:scfringe})
vanishes at $\xi=0$ and $\xi=\imath$. The effect of these poles
is to introduce corners in the map; the angle of these is controlled
by the strength of the pole, here $1/2$. The factor of $\exp(-\pi\xi/2$)
is required to ensure that the boundary of the $\Re(w)<0$ part of
the strip remains horizontal while the $\Re(w)>0$ half is mapped
to the vertical boundaries. The integral can be performed analytically,
\begin{equation}
z(w)=\frac{\sqrt{2}}{\pi}e^{-\frac{\pi w}{2}}\left(1+\frac{e^{\pi w}\arcsin\left(e^{\pi w}\right)}{\sqrt{1-e^{2\pi w}}}\right)\sqrt{\sinh(\pi w)},
\end{equation}
where the overall prefactor is chosen to ensure that the $\Re(w)<0$
half of the strip is mapped onto $0\leq\Im(z)\leq1$. 

The magnetic field strength can be determined using this map as follows:
in the space between the pole pieces $\nabla\cdot\mathbf{B}=0$ and
hence the magnetic field can be constructed from a scalar potential
$\mathbf{B}=-\nabla\psi$ subject to Laplace's equation $\nabla^{2}\psi=0$.
On the boundary, $\mathbf{B}$ must be perpendicular to the pole pieces
so these must be surfaces of constant $\psi$. In the $w$-plane,
the system resembles a parallel plate capacitor with the solution
$\psi=Av+B$ where $A$ and $B$ are arbitrary constants. Since Laplace's
equation is invariant under the conformal mapping, this solution remains
valid when projected into the $z$-plane: the lines of constant $v$
are equipotentials and lines of constant $u$ are the field lines
as shown in figure \ref{fig:CMFringe}.

The magnetic field can be found from $\mathbf{B}=-A\left(\partial_{x}v,\partial_{y}v\right)$
; this can be expressed in complex form in the $w$-plane as $B(w)=-A/z'(w)$\citep{visualcomplexanalysis}.
Evaluating $z(w)$ and $B(w)$ along the center of the strip $w=t+i/2$
yields $B_{y}^{0}$ along the midplane; the $x$ derivative of $B_{y}$
required for the pendulum motion is evaluated from these using the
chain rule,
\begin{eqnarray}
x & = & \frac{1}{\pi}\left[e^{\pi t}\sqrt{1+e^{-2\pi t}}-2\text{arccosh}(e^{\pi t})\right]\nonumber \\
B_{y}^{0} & = & \left(1+e^{2\pi t}\right)^{-1/2}\nonumber \\
\frac{\partial B_{y}^{0}}{\partial x} & = & -\frac{1}{4}\pi\text{sech}(\pi t)^{2}.\label{eq:fringefield}
\end{eqnarray}

Having tabulated values of \textbf{$B_{y}'(x)$ }using the above implicit
expressions, a final form for $\zeta(x)$ was assembled from two appropriately
translated and rescaled copies of $B_{y}'(x)$ as shown in the inset
of figure \ref{fig:CMFringe}; this was used in subsequent calculations
as an \texttt{InterpolatingFunction} in \emph{Mathematica}, or exported
as a table for use in \texttt{FlexPDE}. For the plates and pole pieces
considered, it is actually the case that the plates were slightly
larger than the width of the pole pieces; while this could easily
be accommodated by a simple rescaling of the solution (\ref{eq:fringefield}),
it was found in practice not to affect the dissipation enough to affect
the quality of the fit.

\subsection{Conformal Mapping Solution\label{sub:Conformal-Mapping-Solution}}

\begin{figure}
\begin{centering}
\includegraphics[width=1\columnwidth]{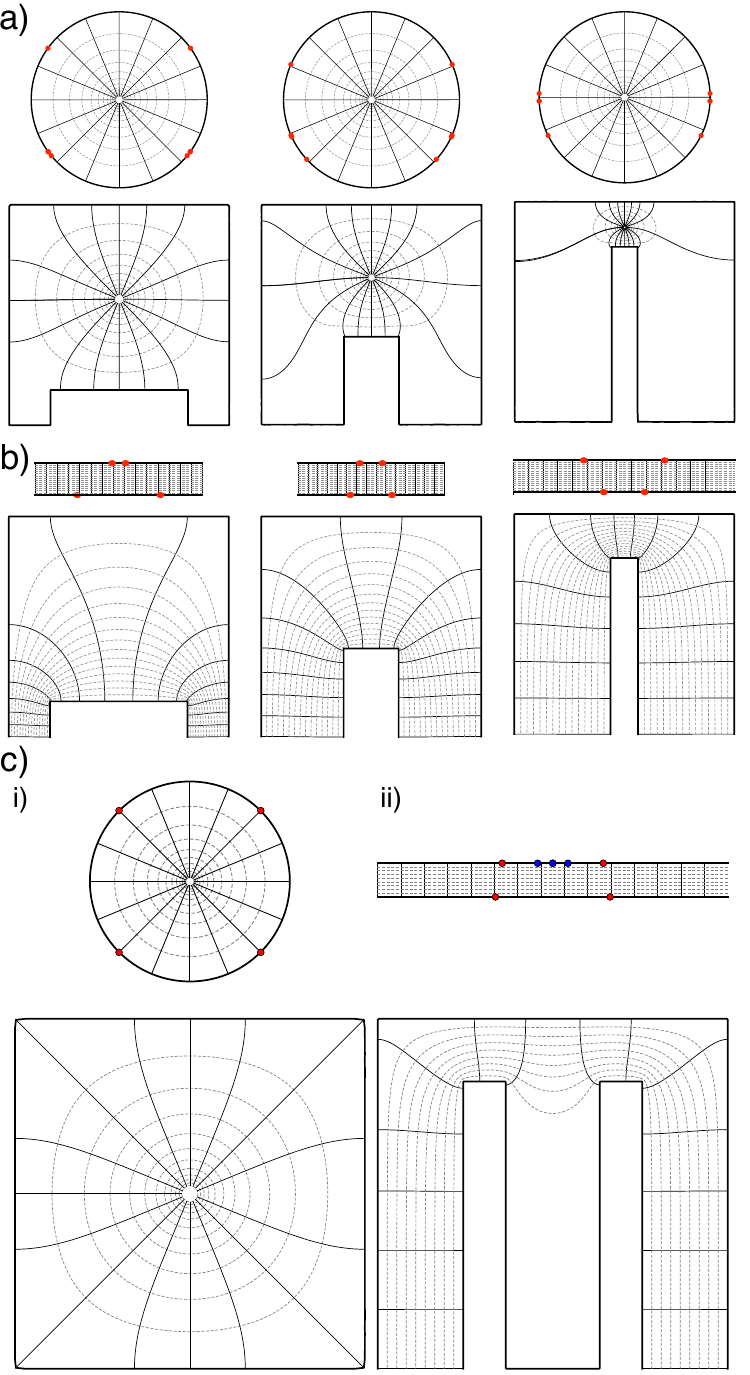}
\par\end{centering}

\protect\caption{\label{fig:ConformalMapping-2}(a) Map from the unit circle to a paddle
with a single slit of variable geometry. (b) Map from a rectangular
domain to the slit. The position of poles in the canonical domains
is indicated with red circles; these become vertices in the image
domain. (c) Two maps relevant to the experiment: i) map from a circle
to a square; ii) map from a rectangular domain to a square with two
slits\textemdash note significant crowding in the central spar. }
\end{figure}

Conformal mapping can be also used to evaluate the power dissipation
of the plate as it passes through the fringing field. The strategy
here is to find a mapping from a simple canonical domain on which
the Green function is known to one that approximates the shape of
the plates; then the integral (\ref{eq:powerdissipation-2}) can be
computed. As will be seen, the choice of canonical domain is important
to render the problem tractable. We therefore examined two choices:
first, the mapping from the unit circle ($w$-plane) to a polygon
of interest ($z$-plane) defined by an ordered set of vertices $z_{i}^{0}$
is given by the Schwarz-Christoffel formula\citep{driscoll2002schwarz},
\begin{equation}
z=A+C\int^{w}\prod_{i}(\xi-w_{i}^{0})^{-\alpha_{i}}\ d\xi\label{eq:sc}
\end{equation}
where $\alpha_{i}$ is the exterior turning angle of the polygon at
vertex $i$ and the $w_{i}^{0}$ are the points on the unit circle
that are mapped to the vertices of the polygon $z_{i}^{0}$. The constants
$A$ and $C$ are translation and scaling factors respectively. While
the set of parameters $\{\alpha_{i}\}$ can be determined directly
from the polygon, the positions of the parameters $\{w_{i}^{0}\}$
must be determined numerically\citep{driscoll2002schwarz}. $A$ and
$C$ are then determined to rotate and scale the result onto the polygon
of interest. Some representative results for a single slit are shown
in fig. \ref{fig:ConformalMapping-2}(a). Due to the symmetry of the
problem, the position of three poles must be determined numerically
with the remaining vertices provided by reflection. 

The second choice of canonical domain is a rectangular portion of
the semi-infinite strip, $-W<\Re(w)<W$ , $0<\Im(w)<1$. A mapping
onto the shape of interest is given by,
\begin{equation}
z=A+C\int^{w}\prod_{i}\left(\sinh\left[\frac{\pi}{2}\left(w-w_{i}^{0}\right)\right]\right)^{-\alpha_{i}}\ d\xi\label{eq:scstripmap}
\end{equation}
where similarly $w_{i}^{0}$ are the position of the poles which produce
the required corners in the image domain and $\alpha_{i}$ are the
turning angles; this formula is a more general version of equation
(\ref{eq:scfringe}) used to determine the fringing field. Results
for the single slit are shown in fig. \ref{fig:ConformalMapping-2}(b).
Note that this map is approximate only: the left and right sides of
the rectangle are mapped to the bottom of the paddle\textemdash the
side that the slit is on\textemdash but their image under (\ref{eq:scstripmap})
is not a straight line. Hence, only two poles $\{w_{i}^{0}\}$\textemdash mapped
to the interior vertices of the slit\textemdash need be determined
numerically; the third parameter is the width of the rectangle $W$.
Despite the approximate nature of the map, the distortion of the lower
boundary is very small. These parameters were found by minimizing
an objective function that was constructed from the $L_{2}$ norm
of the difference between the ratios of the sides of the polygon and
their desired values; a notebook to do so is provided as Supplementary
Material. Robust numerical codes including \texttt{\textsc{SCPACK\citep{trefethen1980numerical}}}
and the SC Toolbox\citep{Driscoll96algorithm756:} are available for
more complex shapes. 

Comparing the two choices of canonical domain in figure \ref{fig:ConformalMapping-2},
an important difference is apparent. The poles in the maps from the
unit circle tend to lie very close to one another, becoming in some
cases indistinguishable in the figure, though they remain numerically
distinct. As a consequence of this phenomenon, known as \emph{crowding}\citep{driscoll2002schwarz},
very small regions adjacent to these crowded poles are mapped to large
regions in the image domain. Crowding is problematic, because it makes
numerically finding the $w_{i}^{0}$ challenging; it will also cause
problems in evaluating the power dissipation. Notice that using the
rectangular source domain eliminates the crowding problem as the poles
remain well-separated. 

In fig. \ref{fig:ConformalMapping-2}(c), we show two maps relevant
to the experiment. First, the map from the circle to the square is
famously
\begin{equation}
z=A+C\int^{w}\frac{1}{\sqrt{1+\xi^{4}}}\ d\xi.\label{eq:squaremap}
\end{equation}
An example with two slits that closely matches the experimental plate
is shown in fig. \ref{fig:ConformalMapping-2}(c)(ii), constructed
by introducing additional poles into one side of the one-slit rectangle
maps above. Unfortunately, the crowding phenomenon is visible in this
map: a very small piece of the rectangle is mapped to the central
spar of the plate. As will be seen later, this will reduce the accuracy
of the power dissipation calculation when this spar coincides with
the edge of the pole pieces. While it is possible to find maps with
crowding localized to different portions of the plate\citep{driscoll2002schwarz},
it cannot be alleviated entirely if more than one slit is present.
Hence, we did not pursue maps for plates with more than 2 slits. 

Turning, then, to the issue of evaluating the power dissipated by
the induced current, we initially tried to use the eigenfunction expansion
of the Green function,
\begin{equation}
G(\mathbf{x}',\mathbf{x}'')=\sum_{n}\frac{\psi_{n}(\mathbf{x})\psi_{n}(\mathbf{x}')}{\lambda_{n}},\label{eq:eigenfunctionexpansion}
\end{equation}
which, inserted into the integral in (\ref{eq:powerdissipation-2})
allows one to obtain, 
\begin{eqnarray}
\int\int\zeta(\mathbf{x}',q)\zeta(\mathbf{x}'',q)G(\mathbf{x}',\mathbf{x}'')\ dA'\ dA''\label{eq:ci1}\\
=\sum_{n}\frac{\left[\int\zeta(\mathbf{x},q)\psi_{n}(\mathbf{x})\ dA\right]^{2}}{\lambda_{n}}\label{eq:conformalintegral}
\end{eqnarray}
The 2D integral in (\ref{eq:ci1}) can then be performed in the $w$-plane
using the normalized eigenfunctions of the unit disk and the mapping
$z(w)$,
\begin{equation}
\int\zeta(\mathbf{x})\psi_{n}(\mathbf{x})\ dA=\int\zeta(z(w),q)\psi_{n}(w)\left|z'(w)\right|^{2}\ dw.\label{eq:conformal2d}
\end{equation}
Unfortunately, we found that the series (\ref{eq:conformalintegral})
converges very slowly except for the square map (\ref{eq:squaremap}).
The problem is further compounded for the unit circle by the fact
that the integrand in (\ref{eq:conformal2d}) is very sharply peaked
and localized to the boundary in many cases due to crowding. For these
reasons, we abandoned this approach and the unit circle maps as viable
means to evaluate the dissipation.

The power dissipation integral (\ref{eq:powerdissipation-2}) can
be re-expressed in the $w$-plane as,

\begin{widetext}

\begin{equation}
D\propto\int\int\zeta(z(w'),q)\zeta(z(w''),q)G(w',w'')\left|z'(w')\right|^{2}\left|z'(w'')\right|^{2}\ dw'\ dw''.\label{eq:dissipation4d}
\end{equation}
\end{widetext}In order to evaluate this integral, one requires a
numerically efficient representation of the Green function. Unfortunately,
the Green's function for a rectangle cannot be expressed in closed
form, requiring expensive evaluation of elliptic integrals. Recently,
however, renewed interest in Green function methods has yielded rapidly
converging series representations\citep{Melnikov2006774}. Briefly,
the strategy is to start from the eigenfunction expansion of $G(w,w')$
(\ref{eq:eigenfunctionexpansion}) on the rectangle and separate the
double summation into two pieces: one that can be summed analytically
and a second that remains to be performed numerically. Fortuitously,
the analytical part contains the inherent singularity; the remaining
numerical sum is a smooth function requiring few terms to converge
to machine precision. Equation (7) of \citep{Melnikov2006774} presents
an expression for the green function in a square; we derived an equivalent
expression for the rectangle $-W<\Re(w)<W$ , $0<\Im(w)<1$ of interest,

\begin{widetext}
\begin{equation}
G(w,w')=\frac{1}{2\pi}\ln\frac{\left|1-e^{w-\bar{w'}}\right|\left|1-e^{w+\bar{w'}}\right|}{\left|1-e^{w-w'}\right|\left|1-e^{w+w'}\right|}-\frac{2}{\pi}\sum_{n=1}^{\infty}\frac{\sinh\left(n\pi u\right)\sinh\left(n\pi u'\right)}{ne^{2n\pi W}\sinh\left(2n\pi W\right)}\sin\left(n\pi v\right)\sin\left(n\pi v'\right)\label{eq:rapidgreenfunction}
\end{equation}
\end{widetext}where $w=u+iv$ and $w'=u'+iv'$ consistent with our
earlier definitions. 

Inserting (\ref{eq:rapidgreenfunction}), (\ref{eq:scstripmap}) and
$\zeta(x)$ derived in the previous section into (\ref{eq:dissipation4d}),
we evaluated the local power dissipation rate $P(q)$ as a function
of $q$ for a variety of plate geometries. A notebook to perform these
calculations is presented as Supplementary Material; we defer consideration
of the results until the subsequent section, in which the same function
is calculated by finite element simulations.

\subsection{Finite Element Simulations\label{sub:Finite-Element-Simulations}}

\begin{figure}
\begin{centering}
\includegraphics[width=1\columnwidth]{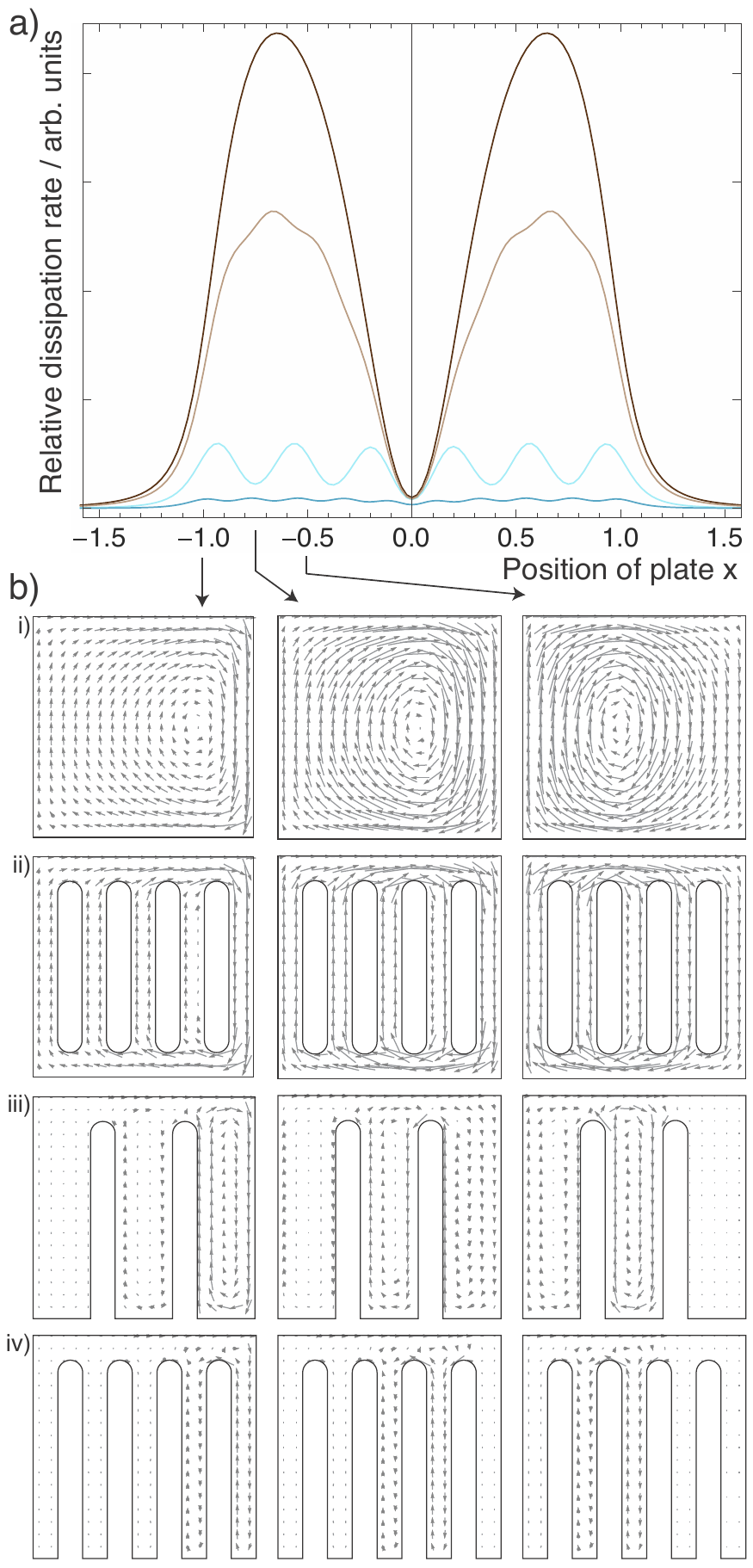}
\par\end{centering}

\protect\caption{\label{fig:Flexpde}(a) Dissipation rate as a function of position
and (b) i-iv corresponding current density profiles obtained numerically
for the various shapes from \texttt{FlexPDE}. }
\end{figure}

Finite element analysis is a commonly used numerical technique for
solving partial differential equations in systems too complicated
for analytical solution. Here, the PDE of interest is, as shown in
the previous sections, Poisson's equation with a spatially dependent
source term (\ref{eq:Poisson}). This must be solved to obtain the
stream function $\phi(x',y')$ for each of the paddle shapes of interest
and at each point $q$ in the motion of the paddle. Suitable input
files for \texttt{FlexPDE} for each plate are included as Supplementary
Material. Each script contains field variable and parameter definitions,
a definition of the PDE, a description of the appropriate domain boundary
and specification of the boundary conditions, and a list of the outputs
desired, i.e. $\phi$, $\mathbf{J}$ and the value of the integral
(\ref{eq:powerdissipation-2}). A relative error of $<10^{-4}$ in
the absolute $\phi$ was requested and used by \texttt{FlexPDE} to
guide adaptive refinement. 

For each plate, the boundary condition on the exterior is simply $\phi=0$,
which forces the stream function to lie tangentially to the edge of
the plate. The plate with holes requires special treatment: around
each of the holes $\phi$ should be constant, i.e. the hole should
be an equipotential surface, but the correct value is not \emph{a
priori} known. A way to obtain the correct solution is to exploit
an electrostatic analogy, modifying Poisson's equation by introducing
a permittivity $\epsilon$, 
\begin{equation}
\nabla\cdot(\epsilon\nabla\phi)=\zeta(x',q).\label{eq:modifiedpoisson}
\end{equation}
The computational domain is extended so that $\phi$ is also defined
inside the holes. Different values of $\epsilon$ are used for the
interior of the plate $\epsilon_{plate}$ and the holes $\epsilon_{holes}$.
As the ratio $\epsilon_{holes}/\epsilon_{plate}\to\infty$, the solution
converges to the desired solution where $\phi$ constant inside the
holes. For the purposes of calculation, $\epsilon_{plate}$ is chosen
arbitrarily to be $1$ and $\epsilon_{holes}$ is successively increased;
the value of (\ref{eq:powerdissipation-2}) converged to a relative
error of $10^{-3}$ for $\epsilon_{holes}>10^{4}$. Plots of the stream
function $\phi$ and the corresponding current densities $\mathbf{J}$
resulting from the calculations are displayed in figure \ref{fig:Flexpde},
together with a plot of the dissipation rate $P(q)$. 

\begin{figure}
\begin{centering}
\includegraphics{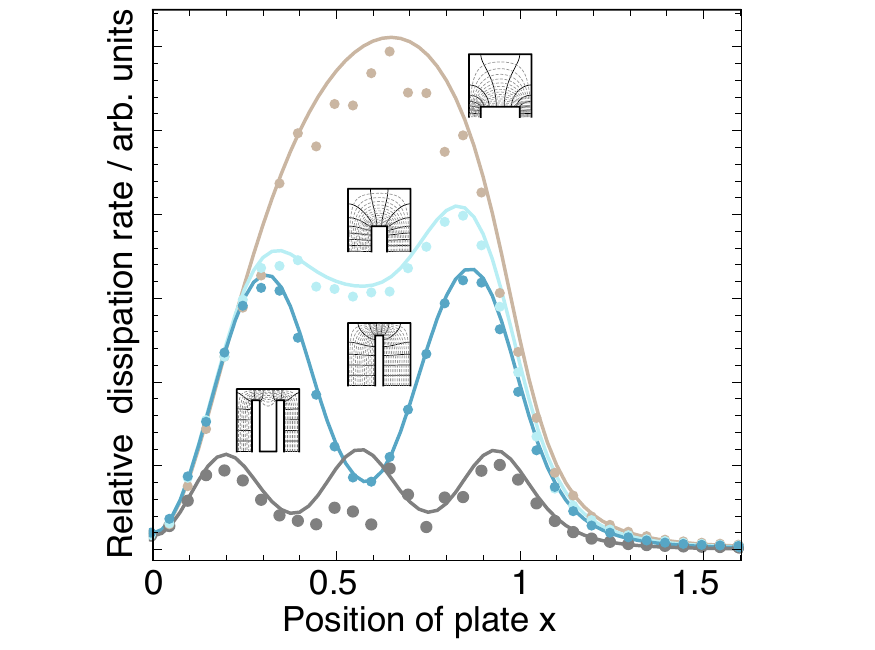}
\par\end{centering}

\protect\caption{\label{fig:FlexPDEcomparison}Comparison of finite element (solid
lines) and conformal mapping (points) approaches to calculating the
local rate of dissipation $P(q)$. }

\end{figure}

It is now possible to compare the Conformal Mapping and Finite Element
methods for this problem. To do so, we used both methods to calculate
the dissipation rate $P(q)$ for the plate geometries displayed in
fig. \ref{fig:ConformalMapping-2}(b) and (c), i.e. those for which
we were able to find appropropriate mappings. To perform the integral
(\ref{eq:dissipation4d}), we used \emph{Mathematica's} \texttt{NIntegrate}
with the \texttt{AdaptiveMonteCarlo} method and $10^{5}$ evaluation
points. The calculation for each plate took $\sim5$ minutes for \texttt{FlexPDE}
and $\sim20$ minutes for the conformal mapping approach on the same
computer (Apple Macbook Air 11'' Late 2014). Results for the various
plates are shown in fig. (\ref{fig:FlexPDEcomparison}). The agreement
is very good for the single slit geometries, with the position and
size of extrema reproduced well by both methods. For the two slit
geometry, however, the conformal mapping method performs less well:
the outer maxima are reproduced well, but the central one is not.
This corresponds to the positioning of the plate where the edge of
the pole piece coincides the central spar, where as discussed in section
\ref{sub:Conformal-Mapping-Solution} the mapping is crowded. Clearly
from this comparison, the method of choice appears to be finite elements,
since the results are less noisy and all plate shapes can be simulated
using the method.

\subsection{Effect on the Motion of the Pendulum\label{sub:Effect-on-the}}

Having formulated both models, we turn to the effect of the dissipation
on the motion of the pendulum. For both the simplified wire model
of section \ref{sec:Simplified-Model} and the magneto-quasistatic
model of section \ref{sec:Magneto-quasistatic-Model}, the equation
of motion has the form,

\begin{equation}
\ddot{x}+\omega_{0}^{2}x+\lambda(x)\dot{x}=0.\label{eq:generalequationofmotion}
\end{equation}
For the wire model,
\begin{equation}
\lambda_{wire}=2\left[\beta+\left(\frac{A^{2}}{L}\right)\frac{B_{0}^{2}}{\rho}\right]
\end{equation}
 as read off from eq. (\ref{eq:equationofmotion}); for the magneto-quasistatic
model the Stokes drag is supplemented by the position-dependent dissipation
rate, 
\begin{equation}
\lambda_{mqs}=2\beta+d\sigma P(q)B_{0}^{2}.
\end{equation}

As is well-known, the equation of motion (\ref{eq:generalequationofmotion})
with constant $\lambda$ can be solved analytically yielding,
\begin{equation}
x(t)=x_{0}e^{-\lambda t/2}\cos(\omega't+\phi)+x_{1}\label{eq:solutionlinear}
\end{equation}
where $x_{0}$ and $x_{1}$ are amplitude and offset parameters, $\phi$
is a phase and $\omega'=\sqrt{\omega^{2}-\lambda^{2}}$. Hence, the
motion remains oscillatory for $\omega>\lambda$. If the plate is
centered with respect to the pole pieces, $x_{1}=0$; $q_{0}$ and
$\phi$ similarly represent initial conditions that are experimentally
controllable. Rather than fit to the entire trajectory, it is convenient
to look only at the extreme points where the pendulum is stationary
and $\cos(\omega_{0}t+\phi)=\pm1$. Inserting this into (\ref{eq:solutionlinear})
and rearranging, we obtain 
\begin{equation}
\log|x(t)|=\log x_{0}-\lambda t/2,
\end{equation}
showing that the values of the position of the pendulum at successive
stationary points should obey a linear relation on a plot of $\log(x)$
versus $t$. 

For the magneto-quasistatic model, analytical solution of (\ref{eq:generalequationofmotion})
is impossible and hence the equation must be solved numerically. This
was performed in \emph{Mathematica}. Because the strategy of looking
at successive extrema on log scales remains a valuable fitting strategy
even if the equation of motion is solved numerically, we additionally
incorporated an extremum-finding routine to identify the stationary
points from the calculated trajectory. With these simplified analytical
and numerical magneto-quasistatic models, we are now ready to compare
them with experimental data.

\section{Experiment}

\begin{figure}
\begin{centering}
\includegraphics[width=3.3in]{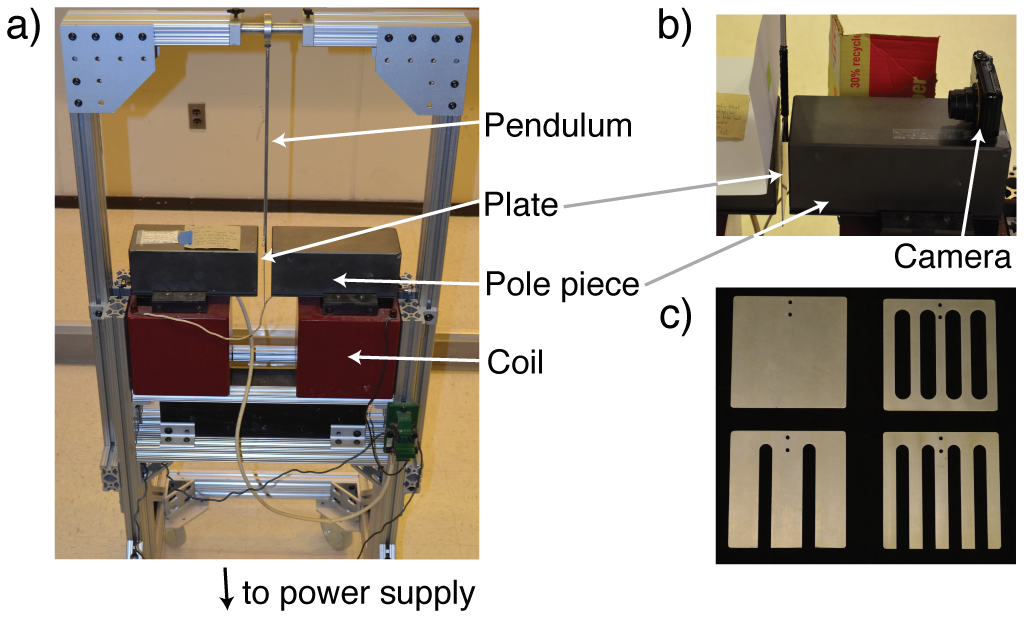}
\par\end{centering}

\protect\caption{\label{fig:App}(a) Photograph of the experimental setup, designed
as a classroom demonstration for introductory physics. (b) Detail
showing placement of camera and screen for imaging. (c) Photograph
of paddle shapes used. }
\end{figure}

In order to test the predictions of the calculations performed in
previous sections, we performed an experiment to observe the motion
of a pendulum with the different paddles attached and fit the models
described in previous sections to the trajectories obtained. 

The magnetic brake apparatus used, shown in Figure \ref{fig:App},
was designed as a classroom demonstration for introductory physics
classes. The apparatus consists of an electromagnet with flat pole
pieces of adjustable separation, powered by a 12V car battery. A potential
divider and ammeter were incorporated into the circuit so that the
current through the magnet could be changed and measured. The pole
pieces are $10\times10$cm and for the present experiment were separated
by $2.5$cm. The pendulum hangs from a supporting frame above the
pole pieces and is made from a solid rod of length 39.5 cm to which
different paddles of interest can be affixed with screws. Four plates
were studied, each of side 11.43 cm by 11.43 cm with different arrangements
of holes or slits as shown earlier in figure \ref{fig:schematic}(d). 

\begin{figure}
\includegraphics{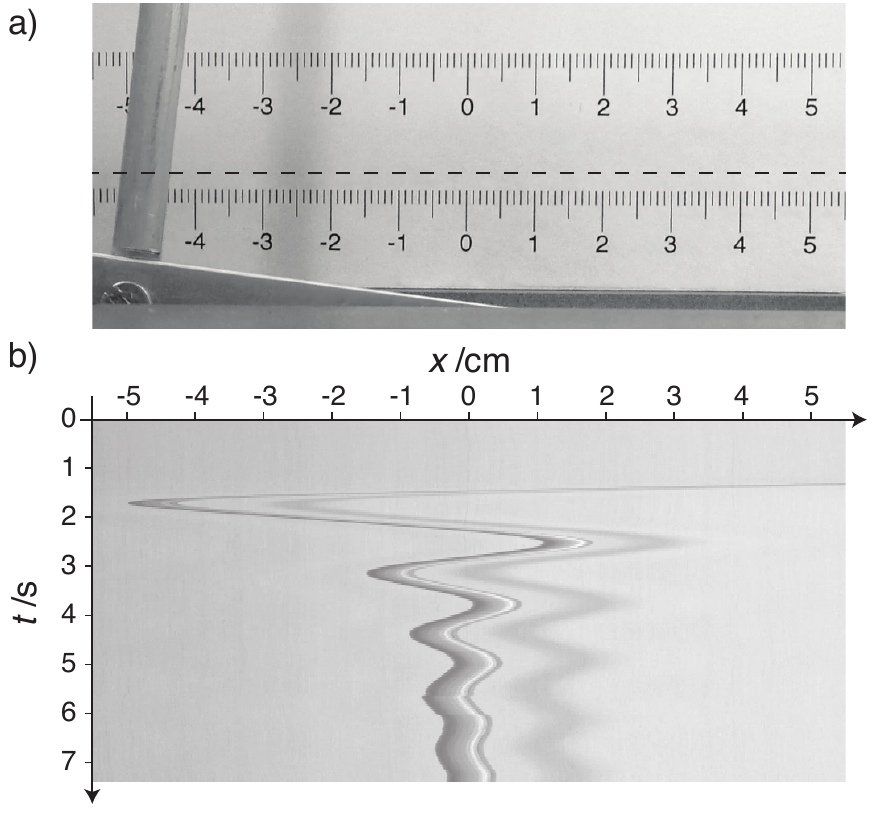}

\protect\caption{\label{fig:Reslice}(a) Sample frames from a video of the pendulum's
motion. From this movie, the row of pixels indicated with the dashed
line is taken from each frame and composited to produce the resliced
image (b). }
\end{figure}

Students collected videos of the motion of the pendulum using a cellphone
camera, and in a later iteration a digital camera was used to achieve
higher image quality and frame rate. Although data logging hardware
could also be used where available, the video approach was adopted
as it is inexpensive, required no additional apparatus beyond the
existing demonstration and is easily reproduced by others. A white
piece of paper with a scale bar was attached to the pole piece behind
the camera to provide a uniform background and facilitate quantitative
analysis of the resulting videos. The camera was positioned so that
the lens was aligned with the rod at rest; its position was marked
with chalk so that it could be easily replaced after each trial. The
plates were released from the same point outside of the magnet, determined
by a ruler placed on top of the pole pieces. The plates were released
when the camera started recording and data was collected until the
pendulum came to a rest. Two sample frames from one of the movies
are shown in figure \ref{fig:Reslice}(a). For each plate, 5 videos
were collected with $0A$, $0.6A$, $1A$, $1.4A$ and $1.8A$ current
flowing through the circuit respectively. Selected videos were taken
more than once to verify repeatability. These values obviously depend
on the details of the circuit, but were chosen as follows: for the
lower bound, a value was sought such that there was a visible difference
between the motion of the plate with weakest damping (the one with
4 slits) in the on and off states. For the upper bound, we found a
current level such that the plate with the strongest damping gave
2-3 oscillations before coming to a stop, i.e. we are in the oscillatory
regime of $\omega>\lambda$ in equation (\ref{eq:generalequationofmotion}). 

Having collected the dataset of videos of the pendulum with different
plates, the trajectories were extracted from the videos. This was
done in two steps: First, the open-source image processing program
\texttt{ImageJ} \citep{imagej} was used to ``reslice'' the movie,
converting it from a stack of images of the $(x,y)$ plane at different
time points $t$ to a new stack of images of $(x,t)$ at different
vertical positions $y$. An sample output is displayed in fig. \ref{fig:Reslice}(b),
showing the damped motion of the pendulum as a function of time for
the row of the movie indicated by the dashed line in fig. \ref{fig:Reslice}(a).
The second step in the data extraction was performed by a custom program
written in \emph{Mathematica}, included as Supplementary Material.
This first subtracted the background, then for each time point located
the position of the bob by finding the centroid of the pixels with
intensity above a certain value. The result of this process was to
yield for each movie a list of $(x,t)$ representing the trajectory
of the pendulum. Using the rulers and known frame rate of the camera
(30fps), the list was calibrated to physical coordinates; the program
also subtracts the equilibrium position of the pendulum from the $x$
position. To facilitate comparison with the models, the extremum-finding
routine mentioned in the previous section was used to identify a list
of stationary points $\{(x_{i},t_{i})\}$ from each trajectory. A
constant was subtracted from the times such that for each trajectory
the first extremum was set to $t=0$.

\section{Model Comparison}

\begin{figure*}
\begin{centering}
\includegraphics{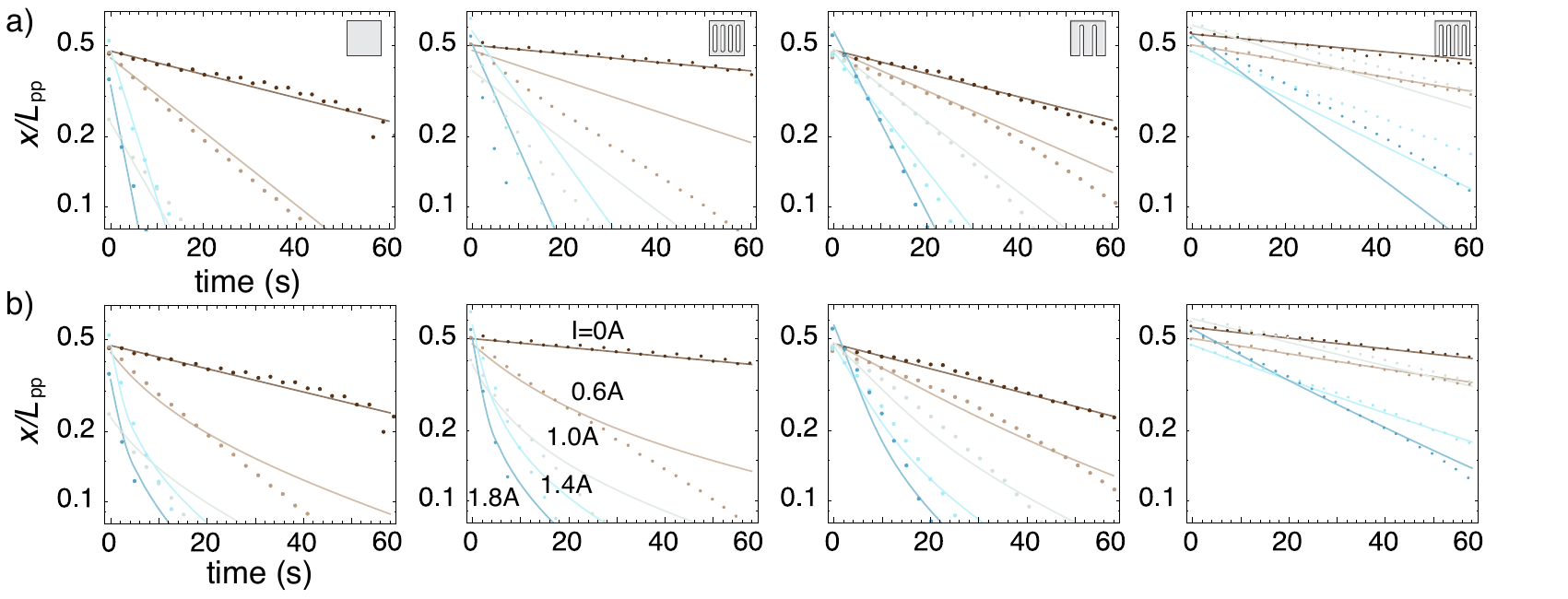}
\par\end{centering}

\protect\caption{\label{fig:Fit}Fits for (a) the simplified wire model and (b) the
magneto-quasistatic model. Experimental data is shown as points; the
model is shown as lines.}
\end{figure*}

Having obtained a set of experimental trajectories, we now fit both
models to the data. Beginning with the simplified wire model, the
analytical solution in section \ref{sub:Effect-on-the} implies that
for each individual trajectory the extrema should follow a straight
line on semi-log scales with slope
\begin{equation}
\lambda_{wire}=\beta+\left(\frac{A^{2}}{L}\right)\frac{B_{0}^{2}}{\rho}.
\end{equation}
From the set of results displayed as points in fig. \ref{fig:Fit},
it is seen that this is generally the case, though some deviations
are apparent. In particular, the motion starts to decay more rapidly
once $\left|x/L_{pp}\right|\lesssim0.1$; we therefore excluded this
data from our fit. For each plate, several values of current $I$
were used to take a succession of trajectories. Although it is obvious
that $B_{0}\propto I$, we did not explicitly measure the magnetic
field and therefore absorb this unknown constant of proportionality
together with $\rho$ the effective resistivity of the wire into a
single fitting parameter $\alpha$. This fitting parameter could,
of course, in principle be calculated or measured separately, but
the purpose of fitting here is to establish whether the model is consistent
with the results for all plates simultaneously. The slopes extracted
from the ensemble of plates with different $\Gamma=A^{2}/L$ (calculated
values are shown in table \ref{tab:FitSimplified}) at different applied
currents $I$ should therefore universally be described by, 
\begin{equation}
\lambda_{wire}=\beta+\alpha\Gamma I^{2}.
\end{equation}
Estimates for the initial position of each trajectory $x_{0}$ were
obtained by fitting a straight line to the extrema individually on
semi-log scales. We then estimated $\beta$ for each plate from the
$I=0$ trajectory for each plate; the results are shown in table \ref{tab:FitSimplified}.
Two pairs of plates seem to cluster around similar values; these were
collected at different times indicating that the natural damping of
the apparatus, perhaps lubricant in the hinge, changed between them. 

\begin{table}
\begin{centering}
\begin{tabular}{cccc}
\hline 
 & $\Gamma/\Gamma_{square}$ & $\beta$ & $\alpha$\tabularnewline
\hline 
\hline 
Square & 1 & $1.1\times10^{-2}$ & 0.060\tabularnewline
4 holes & 0.441 & $4.4\times10^{-3}$ & 0.130\tabularnewline
2 slits & 0.348 & $1.2\times10^{-2}$ & 0.074\tabularnewline
4 slits & 0.136 & $5.2\times10^{-3}$ & 0.044\tabularnewline
\hline 
\end{tabular}
\par\end{centering}

\protect\caption{\label{tab:FitSimplified}Geometric parameters and fitted $\alpha$,
$\beta$ for each plate using the simplified model.}
\end{table}

Results from the universal least-squares one-parameter fit for $\alpha$
are shown in fig. \ref{fig:Fit}(a). As can be seen, the fit isn't
bad and follows many of the trajectories quite well. That said, there
are many discrepancies. We therefore estimated $\alpha$ from each
individual plate using the data at different applied field. While
the details of these are not shown here, the quality of fit was very
much better. The resultant values of $\alpha$ are displayed in table
\ref{tab:FitSimplified}. In particular, it is not surprising that
the plate with 4 holes is such an outlier as the wire model cannot
account for this geometry. Values for the other plates are the same
order of magnitude, which together with fig. \ref{fig:Fit}(a) demonstrates
that the simplified model gives reasonable estimates for the influence
of plate geometry even though it is based on quite egregious approximations. 

In the same spirit, we assume that the damping term in (\ref{eq:generalequationofmotion})
for the magneto-quasistatic model can similarly be described by a
universal constant $\alpha$,
\[
\lambda_{mqs}(x)=\beta+\alpha P(x)I^{2}.
\]
For this model, as well as $x_{0}$, $\beta$, $\alpha$, $P(x)$
and $I$, it is also necessary to specify the natural angular frequency
$\omega$; the value from the $I=0$ trajectory for each plate is
used since $\beta\ll\omega$ for the weak damping in the absence of
the magnetic field and hence $\omega'\approx\omega$. The results
of the one-parameter universal fit for $\alpha$ are displayed in
\ref{fig:Fit}(b). The quality of fit is significantly better, especially
for large amplitudes and early times, and more consistent between
plates than for the simple model, implying that the magneto-quasistatic
model yields good predictions of the effect of plate geometry. Of
particular importance is the fact that the fit for the 4 hole plate
is no worse than that of the other plates, suggesting that these features
have been handled correctly by the model. 

Nonetheless, there remains an unexplained phenomenon in the data,
i.e the anomalous increase in damping once the amplitude of the motion
drops below $x/L_{pp}\sim0.1$. This is not predicted by either model.
The simplified model predicts a constant rate of damping, while the
magneto-quasistatic model predicts a crossover from magnetically dominated
damping at large amplitudes to Stokes dominated damping at small amplitudes.
We tried adjusting the magneto-quasistatic model in various ways\textemdash offsetting
the position of the plate with respect to the pole pieces and changing
the size of the plate relative to the pole pieces\textemdash but found
no similar effect for any reasonable parameters for the experiment.
Clearly, the Stokes form of the non-magnetic damping is most likely
at fault, because the source of damping is likely to be primarily
the hinge rather than air resistance; further analysis should be performed
to determine the origin of this anomaly.

\section{Pedagogical Context}

The scientific results presented in this paper were obtained in part
by students in Spring 2013 as part of the Electromagnetic Theory II
class at Tufts University, the second course in the E\&M sequence
in the graduate program at Tufts; two students from the class are
co-authors on the paper (CW and JW) having spent a considerable time
subsequently acquiring additional data and performing calculations.
In this section, the pedagogical context of these activities is briefly
documented in the hope that it is of use to others. The idea of incorporating
(simple) experiments in class was motivated by an observation that
key scientific skills expected of graduate students\textemdash connecting
theory and experiment; article reading; journal selection and article
preparation\textemdash are not included explicitly in the curriculum.
The project was spread over the second half of the semester as follows:\textemdash{}
\begin{enumerate}
\item A prepatory homework where students derived the magneto-quasistatic
model \ref{sec:Magneto-quasistatic-Model}. 
\item Two $1\frac{1}{4}$ hour class periods where students worked in groups.
The students were divided into groups of four and worked on: i) conformal
mapping analysis, ii) finite element analysis and iii) experiments.
The instructor provided scaffolding activities to help students learn
FlexPDE and the Schwarz-Christoffel mapping technique. These activities
continued subsequently outside of class. 
\item A reading activity whereby students had to identify possible journals
for publication of this work; this was followed by a 15 minute discussion
of journal selection in class.
\item A Just-In-Time Teaching\citep{Novak:1999:JTB:553365}-like pre-class
activity in which students had to identify the structural elements
in a typical paper and construct an outline; these were passed round
anonymously for peer feedback. 
\item A homework activity in which students prepared an introduction for
the paper. Students were asked to read each other's work and identify
strong features as well as possible modifications as a pre-class activity.
This was followed by a discussion in class. 
\item Each group collaboratively produced a draft section for the final
paper as part of a homework; these were combined by TJA into a coherent
document and extensively rewritten during the second iteration of
experiments. 
\end{enumerate}
The focus of the class and restricted time available necessitated
some design trade-offs. For example, we chose to use an existing commercial
software package, \texttt{FlexPDE}, as opposed to implementing a custom
finite element program, for several reasons: First, the scale and
complexity of problems solvable with a specialist code is much greater
than with a simple custom-written program and hence should be applicable
to situations encountered in student's research projects. Second,
the program contains advanced features such as adaptive refinement\textemdash allocation
of grid points based on estimated error\textemdash that are time-consuming
to implement. Third, by removing the focus of the exercise from programmatic
implementation, time can be spent on understanding how to properly
interpret the output using fundamental principles of numerical analysis
such as order, error estimation and convergence. 

While the project-based approach is well-grounded in a theory of learning
(constructivism), due to the small $N=13$ size of the class, it is
difficult to rigorously demonstrate the effectiveness of the activity.
Future iterations would strongly benefit from pre and post testing.
Even so, the outcomes of the project indicate that students made considerable
progress towards learning these challenging scientific skills. Moreover,
the class was very highly rated in the formal feedback mechanism.
It is hoped that, in future, physics education researchers might attempt
a thorough analysis of this and related approaches at the graduate
level.

\section{Conclusion}

The present work has provided a careful, quantitative analysis of
a familiar classroom demonstration, the van Waltenhofen pendulum.
A magneto-quasistatic model has been shown to successfully predict
the relative damping of a selection of plates with different geometry
using conformal mapping and finite elements as tools to carry out
the necessary calculations. As a by-product, a number of valuable
resources for more general use, including plots of the current distribution
during the course of the motion of the pendulum, have been produced.
Additionally, a greatly simplified model, suitable for use in introductory
physics classes, has been formulated that could be used to make the
classroom demonstration more quantitative.

\section*{Acknowledgements}

\emph{The authors wish to thank students from the Tufts Electromagnetic
Theory II class in Spring 2013, the Tufts Department of Physics \&
Astronomy for providing equipment for the project, and Badel Mbanga
and Chris Burke for helpful discussions. The authors contributed to
the paper as follows: CW, JW and PW obtained the experimental data
used in the paper; conformal mapping solutions were obtained by JW
and TJA; finite element simulations were performed by TJA based on
initial results obtained by students in the class; the paper was written
by TJA from sections submitted by each group and revised collectively. }

\end{document}